\documentclass[%
 reprint,
 amsmath,amssymb,
 aps,
]{revtex4-2}

\usepackage{graphicx}
\usepackage{dcolumn}
\usepackage{dsfont}
\usepackage{amssymb}
\usepackage{comment}
\usepackage{xcolor}
\usepackage{float}
\usepackage{bm}
\usepackage{hyperref}

\usepackage{physics}

\begin{document}

\preprint{APS/123-QED}

\title{Autonomous stabilization of many-body entanglement with Floquet Hamiltonians and weak measurement}

\author{Charlotte Franke}
\email{cf590@cam.ac.uk}
\author{Dorian A. Gangloff}%
 \email{dag50@cam.ac.uk}
\affiliation{%
 Cavendish Laboratory, University of Cambridge, JJ Thomson Laboratory, CB3 0US, United Kingdom
}%

\date{\today}

\begin{abstract}

Reaching a technological advantage with large quantum systems requires safeguarding their many-body entanglement. Dissipation typically acts to decohere a quantum system via random projective noise but, when judiciously engineered together with coherent interactions, it can funnel the system towards a target entangled state. The native interactions and dissipative channels available to most systems are, however, difficult to combine effectively. Here we propose interleaving Floquet Hamiltonian engineering, which allows the construction of non-native coherent interactions, with weak measurement, which enables a tuneable dissipative channel, to enable programmable and autonomous stabilization of many-body entanglement. We show this analytically and numerically for the central-spin system of a semiconductor quantum dot, for which we construct spin-squeezed and Schr\"odinger-cat states that are stabilized against realistic levels of dephasing. Our approach is applicable to any system compatible with periodic drives and tuneable measurement strength and could enable novel approaches to practical error correction.

\end{abstract}

                            
\maketitle

Many-body entangled states are a central resource for quantum advantage, as for example in quantum-enhanced metrology with spin ensembles or quantum error correction with bosonic modes. Gaussian
spin-squeezed states redistribute quantum fluctuations to push measurement
sensitivities beyond the standard quantum
limit~\cite{Kitagawa.1993,Wineland.1994,Pezzè.2018}.
Non-Gaussian states, such as collective Schr\"odinger-cat states (coherent
superpositions of macroscopically distinct spin configurations), exhibit
Wigner negativity and provide a resource for quantum error correction
\cite{Mirrahimi.2014,Leghtas.2015,Lescanne.2020,Franke.2026}.
Preparing either class in a large ensemble is hard: coherent protocols need
precise initialization and sustained high-fidelity control, and are therefore
highly sensitive to decoherence and imperfections.

Dissipative state engineering offers an alternative. By designing how the system
couples to a dissipative reservoir, the target state is stabilized
autonomously, making the preparation robust against initialization errors and
against perturbations weaker than the engineered dissipation. Building on early
proposals~\cite{Poyatos.1996,Verstraete.2009,Torre.2012}, such
schemes have deterministically generated entangled states in trapped
ions~\cite{Barreiro.2011,Lin.2013}, atomic ensembles~\cite{Krauter.2011}, and
superconducting circuits~\cite{Shankar.2013}. In many systems, however, the interactions needed to stabilize the target are
not available natively and must be built through Hamiltonian engineering.
Floquet engineering does this with periodic pulse sequences, generating
effective interactions by time averaging over the coherent
dynamics~\cite{Bukov.2015,Goldman.2014,Geier.2021}. Dissipation, however, is normally continuous, and when present during the drive is not diagonal in the driven basis, disturbing the coherent averaging the drive relies on. One solution is to separate these processes in
space, driving the system while a distinct reservoir cools it into an eigenstate
of the resulting Floquet Hamiltonian~\cite{Petiziol.2022,Ritter.2025, Raman.2026}. Others
instead make use of the measurement record, either feeding it back to steer the
system~\cite{Wiseman.1993,Thomsen.2002,Geremia.2003} or keeping it,
so that the prepared state depends on the sequence of
outcomes~\cite{Lantaño.2025}. Our solution is to combine the two steps stroboscopically: first a coherent Floquet window, in which a pulse sequence builds an effective system interaction, and second a weak
measurement, which imprints the corresponding dissipative
channel onto a subset of the system. Here the measurement record is discarded, and every run relaxes autonomously to the steady state.

We demonstrate our method through the lens of a concrete example: the central-spin system -- a nuclear spin ensemble coupled to a single, optically addressable electron spin, as for example in a quantum dot~\cite{Urbaszek.2013}. Because the electron both couples to the nuclei and is measured, the drive and the dissipation must be separated in time; they cannot act at once without the
dissipation disturbing the coherent evolution. The coherent Floquet window is a pulse sequence on the electron which builds an effective electron-nuclear interaction. The weak measurement is a state-selective optical readout
of the electron with tuneable strength, which imprints the corresponding dissipative
channel on the nuclei. If the electron decoheres rapidly compared to the
engineered coupling it can be adiabatically eliminated~\cite{Rudner.2011}, and over many cycles the coarse-grained dynamics converge to an effective Lindblad master equation for the nuclei, whose dark state is the stabilized target. 

We show the dissipative stabilization of two many-body states, a
spin-squeezed state and a collective Schr\"odinger cat, by simulating the full stroboscopic
dynamics. For the squeezed state we generate an electron-nuclear
flip-flop interaction at zeroth order~\cite{Denning.2019}. The resulting dark state is squeezed more than an order of
magnitude below the standard quantum limit. To stabilize a cat we design a new
pulse sequence whose first-order Floquet term engineers a two-quantum flip-flop,
an interaction not available natively. The jump operator is now quadratic in the
nuclear spin, and its dark state is a collective cat, stabilized
with fidelity above $0.9$ across a range of sizes and developing pronounced
interference fringes. In both cases the target state is protected by a finite Liouvillian gap, and
robust against realistic noise such as collective and local nuclear dephasing.
Our protocols can be implemented on current quantum-dot platforms:
a nuclear coherence of only $T_2\approx1$ms~\cite{Dyte.2025}, reached with a single refocusing
pulse and further extended by longer dynamical decoupling sequences, already stabilizes squeezing a few dB below the standard quantum limit
and cat states at fidelity $\approx0.95$.

\begin{figure}
\centering
\includegraphics[width=\linewidth]{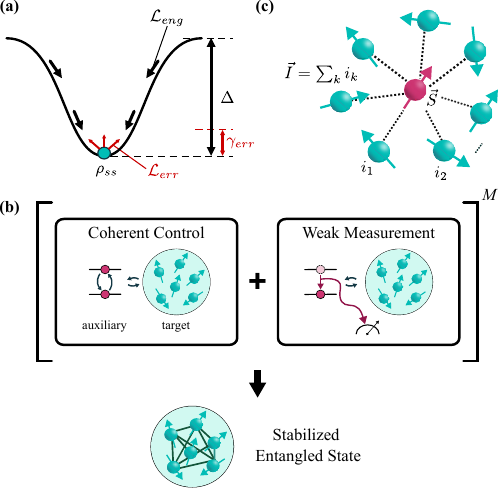}
\caption{\textbf{Stroboscopic dissipative engineering.}
\textbf{(a)} The engineered dissipation $\mathcal{L}_{\rm eng}$ drives the system
into a steady state $\rho_{ss}$, protected by the Liouvillian gap
$\Delta$ against residual error processes $\mathcal{L}_{\rm err}$
acting at rate $\gamma_{\rm err}\ll\Delta$.
\textbf{(b)} Each cycle interleaves a coherent-control window, in which an
auxiliary system interacts unitarily with the target, with a weak
measurement of the auxiliary that induces dissipation. Repeating the
cycle $M$ times stabilizes the target in an entangled steady state.
\textbf{(c)} Central-spin realization: a single electron spin $\vec S$
(auxiliary) is collectively coupled to a nuclear ensemble described by
$\vec I = \sum_k \vec i_k$ (target).}
\label{fig:protocol}
\end{figure}

{\it Stroboscopic dissipative engineering ---} Dissipative state engineering prepares a target state as the steady
state of a controlled Lindblad evolution for the density matrix $\rho$,
\begin{equation}
    \dot{\rho}
    =\mathcal{L}_{\rm eng}\rho
    \equiv
    -i[H,\rho]
    +
    \Gamma\mathcal{D}[L]\rho,
    \label{eq:lindblad}
\end{equation}
where $H$ is the Hamiltonian, $\mathcal{D}[L]\rho = L\rho L^\dagger
- \tfrac{1}{2}\{L^\dagger L,\rho\}$ is the dissipator built from the engineered
jump operator $L$, and $\Gamma$ is its rate. Together, $H$ and $L$ fix an
attractive steady state $\rho_{\rm ss}$ of the engineered Liouvillian
$\mathcal{L}_{\rm eng}$ \cite{Verstraete.2009}. The advantage of this approach is passive protection:
the engineered dynamics continuously drive the system back towards
$\rho_{\rm ss}$, counteracting noise before it can accumulate
[Fig.~\ref{fig:protocol}(a)]. The rate of this relaxation is set by the
Liouvillian gap $\Delta$, the slowest nonzero relaxation rate of
$\mathcal{L}_{\rm eng}$. Any noise process $\mathcal{L}_{\rm err}$ that drives the system out of the steady-state manifold is thus suppressed, at order $\gamma_{\rm err}/\Delta$.

We realize Eq.~\eqref{eq:lindblad} stroboscopically using an auxiliary
system [Fig.~\ref{fig:protocol}(b)]. Each cycle consists of two steps: a
coherent-control window, in which the auxiliary interacts unitarily
with the target under an engineered Hamiltonian $H$, followed by a weak
measurement of the auxiliary that imprints the dissipative channel $L$
on the target.
The measurement strength is set by a small parameter $\epsilon\ll1$, the
per-cycle probability of projecting the auxiliary. Repeating the cycle $M\gg1$ times reproduces Eq.~\eqref{eq:lindblad} as a
coarse-grained description of the dynamics, with dissipation rate
$\Gamma=\epsilon/\tau$ ($\tau$ the cycle duration) and with $H$ and $L$ set
independently by the two halves of the cycle. Because coherent and
dissipative steps are separated in time, dissipation no longer
interferes with pulse-based Hamiltonian engineering. Although each measurement is weak, the measurement-induced decay of the auxiliary
is fast compared with the induced target dynamics, so the auxiliary can be
adiabatically eliminated, leaving an effective master equation for the target
alone (see App.~\ref{app:framework} for the derivation).

\begin{figure*}
  \centering
  \includegraphics[width=\linewidth]{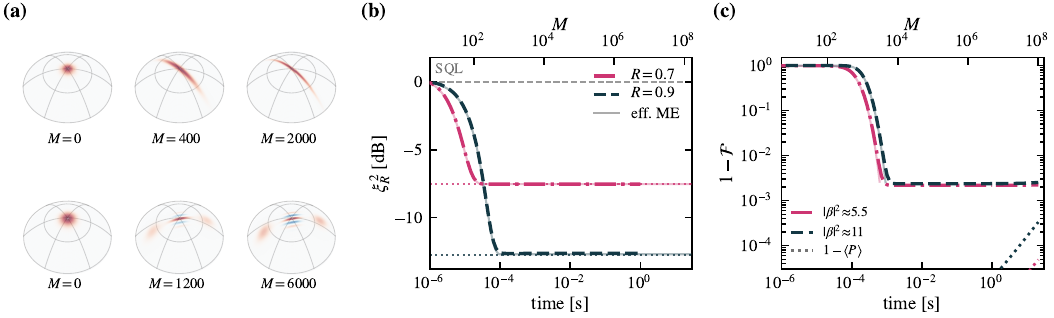}
  \caption{\textbf{Stabilization of the two targets.}
  \textbf{(a)} Wigner snapshots on the Bloch sphere at cycle number $M$: the
  squeezed state (top) contracts along one quadrature, and the
  cat (bottom) splits off the pole into two lobes with interference fringes. Shown on a reduced ensemble ($I=100$) for
  visibility.
  \textbf{(b)} Squeezing parameter $\xi_R^2$ versus time for ratios
  $R=|\Omega_-/\Omega_+|=0.7,0.9$, driven below the SQL (dashed) to the steady
  floors (dotted).
  \textbf{(c)} Cat infidelity $1-\mathcal{F}$ to the target even cat versus time, for sizes
  $|\beta|^2\approx5.5$ and $\approx11$. Dotted: the parity error $1-\langle P\rangle$, which drifts up within the gapless 2D dark manifold and grows with cat size, staying below $10^{-3}$ out to $\sim20$s.
  In (b,c) the top axis is the stroboscopic cycle number $M$ (cycle time $t_c\approx180$~ns for squeezing, $205$~ns for the cat). Solid curves are calculated from the effective master equations. The operating point is $\omega_n/2\pi=100$~MHz, $A_{\rm c}/\omega_n=0.0011$ and $I=10^4$,
  with $A_{\rm nc}/\omega_n=1.35\times10^{-4}$ (squeezing) and $7\times10^{-5}$ (cat).
  The per-cycle measurement is set at the stabilization optimum (App.~\ref{app:HP_Bogo}): $\epsilon\approx0.03$ (squeezing) and $0.03$ (cat), below the weak-measurement cap $\epsilon\le0.1$.}
  \label{fig:ideal}
\end{figure*}

{\it Engineered interactions and target states ---} The central-spin system of a quantum dot~\cite{Urbaszek.2013,Gangloff.2019,Appel.2025}, as shown in Fig.~\ref{fig:protocol}(c), provides a natural platform for
this construction. A single
electron spin $\vec S$ couples to a nuclear ensemble with collective spin
$\vec I = \sum_k \vec I^{(k)}$, where $\vec I^{(k)}$ is the spin of nucleus $k$; we
take the $N$ nuclei to be spin-$\tfrac12$ for simplicity, so the ensemble fully polarized along an external $\hat z$ field has $I_{\max}=N/2$ and energy $\omega_nI_z$; the collective treatment depends only on the total spin
$I_{\max}=I_0N$ and carries over straightforwardly to higher on-site spins $I_0=\{3/2,9/2\}$ of typical quantum-dot
nuclei~\cite{Urbaszek.2013}.
With the electron quantized along $\hat z$, the hyperfine interaction consists
of a collinear (longitudinal) term $A_cS_z I_z$ and a weaker non-collinear
term $A_{\rm nc}S_z I_x$, which couples the electron to a transverse nuclear
component~\cite{Hogele2012a,Shofer.2024}. We describe the ensemble in the collective
(Dicke) basis $\{\lvert I,m\rangle\}$ of $I_z$ eigenstates ($m=-I,\dots,I$),
with the fully polarized state $\lvert I,-I\rangle$ at the south pole of the
collective Bloch sphere. Periodic driving of the electron during the coherent
window converts the non-collinear hyperfine into an effective collective
electron-nuclear exchange, tailored to the state we wish to stabilize. In
terms of the ladder operators $S_\pm = S_x \pm i S_y$ and
$I_\pm = I_x \pm i I_y$, the two target entangled states of this work each require a specific
interaction Hamiltonian:
\begin{align}
  H_{\mathrm{lin}}  &= \Omega_+ (S_- I_+ + S_+ I_-)
                     + \Omega_- (S_+ I_+ + S_- I_-),
  \label{eq:Hlin}\\
  H_{\mathrm{quad}} &= \chi(S_+ I_-^2 + S_- I_+^2) + \Omega S_x.
  \label{eq:Hquad}
\end{align}
The linear exchange Hamiltonian $H_{\mathrm{lin}}$ superposes flip-flop and flip-flip processes; the quadratic exchange Hamiltonian $H_{\mathrm{quad}}$, with
coupling $\chi$, is a single two-quantum flip-flop, transferring a pair of
nuclear quanta per electron flip. Following a Floquet Hamiltonian construction, 
averaging the hyperfine interaction yields the flip-flop and flip-flip processes at zeroth order. Concatenating the two gives the target linear exchange with the weights $\Omega_\pm$ tuneable by their relative frequency in the pulse sequence. The two-quantum exchange arises one order higher, where the commutator of two drive
harmonics squares the nuclear coupling into $I_\pm^2$ -- Appendix ~\ref{app:eng} provides the derivation of both effective Hamiltonians. This construction remains valid in the weak coupling regime $A_{\rm nc}/\omega_n\ll1$.

At the end of each cycle, the electron is weakly measured, relaxing it to $\ket{\downarrow}$ with small probability. Averaged over outcomes, this measurement is the dissipator $\kappa \mathcal{D}[S_-]$. When the measurement rate dominates the
engineered couplings, the electron can be adiabatically eliminated to leave an effective master equation for the nuclei (App.~\ref{app:framework}). Combined with the linear exchange $H_{\mathrm{lin}}$, this gives the master equation
\begin{equation}
  \dot\rho_n = \Gamma_{\mathrm{sq}}\mathcal{D}[L_{\mathrm{sq}}]\rho_n,
  \qquad
  L_{\mathrm{sq}} \propto \Omega_+ I_- + \Omega_- I_+,
  \label{eq:Lsq}
\end{equation}
which damps the ensemble, at rate
$\Gamma_{\mathrm{sq}} = 4(|\Omega_+|^2 + |\Omega_-|^2)/\kappa$, toward the state
that $L_{\mathrm{sq}}$ annihilates. This is the dissipative
squeezing of Ref.~\cite{Torre.2012}, here realized through the engineered
electron-nuclear exchange rather than an optical cavity. The dark state is spin-squeezed,
with the squeezing growing as the two weights approach equality,
$|\Omega_-| \to |\Omega_+|$, at the cost of a vanishing Liouvillian gap and hence
slower convergence.

When weak measurement is combined with the quadratic exchange $H_{\mathrm{quad}}$, we obtain the master equation
\begin{equation}
  \dot\rho_n = \Gamma_{\mathrm{cat}}\mathcal{D}[I_-^2 - \alpha^2]\rho_n,
  \qquad
  \alpha^2 = -\Omega/2\chi,
  \label{eq:cat_me}
\end{equation}
with $\Gamma_{\mathrm{cat}} = 4\chi^2/\kappa$. This two-photon dissipator
stabilizes a cat, as in bosonic modes and collective
spins~\cite{Mirrahimi.2014,Leghtas.2015,Qin.2021,Liu.2025cbo}. Its dark states
satisfy $I_-^2\lvert\psi\rangle = \alpha^2\lvert\psi\rangle$: in the large-$I$
limit these are the two spin-coherent states~\cite{Arecchi.1972}
$\lvert\pm\beta\rangle$ of size $\lvert\beta\rvert^2 = \lvert\alpha^2\rvert/2I = \lvert\Omega\rvert/4I\chi$
(without the drive, $\alpha=0$, the fully polarized pole). Because the jump is
quadratic it cannot distinguish $\lvert\beta\rangle$ from $\lvert{-}\beta\rangle$,
so the stabilized dark state is their superposition: a two-lobe
Schr\"odinger-cat state. Because $I_-^2$ conserves parity, the parity of the initial state selects whether the even or odd cat is stabilized.


{\it Numerical demonstration ---} We validate both protocols by simulating the full stroboscopic dynamics:
coherent evolution under the Floquet-engineered interaction, followed by a weak
measurement on the electron (numerical methods in App.~\ref{app:numerics}). This
tests the whole construction at once: that the
pulse sequence builds the intended electron-nuclear interaction, that
alternating coherent windows and weak measurements gives the target Lindblad
dynamics, and that the electron can be adiabatically eliminated. Each run starts
from the polarized state $|I,-I\rangle$, and we calculate the full dynamics
as well as the dynamics under the effective nuclear master equations; equivalently, the protocol can be run from the opposite pole $|I,+I\rangle$, interchanging $I_\pm$, to stabilize the mirror targets there. We
present the results in two steps: first the stabilization of each target without
added noise and in the ideal Floquet limit ($A_{\rm nc}/\omega_n\ll1$), then their robustness once nuclear dephasing and Floquet break-down are included.


We quantify the squeezed state by the Wineland parameter~\cite{Wineland.1994},
\begin{equation}
  \xi_R^2 = \frac{2I(\Delta I_\perp)^2_{\min}}{|\langle \vec I\rangle|^2},
  \label{eq:wineland}
\end{equation}
where $(\Delta I_\perp)^2_{\min}$ is the smallest variance of the collective
spin along any direction perpendicular to the mean spin $\langle\vec I\rangle$.
It gives the factor by which the metrological phase variance beats
($\xi_R^2<1$) or falls short of ($\xi_R^2>1$) the standard quantum limit (SQL).
We quantify the cat by
the fidelity $\mathcal{F}=\langle\psi_{\rm cat}\rvert\rho_n\lvert\psi_{\rm cat}\rangle$ of
the nuclear state $\rho_n$ to the target even cat
$\lvert\psi_{\rm cat}\rangle=\mathcal{N}(\lvert\beta\rangle+\lvert{-}\beta\rangle)$,
with $\mathcal{N}$ a normalization constant.

Both targets form from the polarized pole under the engineered dissipation
[Fig.~\ref{fig:ideal}(a)]. For the linear jump the transverse fluctuations are
squeezed onto one quadrature and $\xi_R^2$ rapidly drops below the SQL, settling at a
steady value [Fig.~\ref{fig:ideal}(b)] within $10^3$ cycles. The steady-state value depends on the weight ratio
$R=|\Omega_-/\Omega_+|$: as $R\to1$ the squeezing deepens, matching the analytic
floor $\xi_R^2=(1-R)/(1+R)$ obtained by linearizing about the
pole~\cite{Torre.2012}. The cost of deeper squeezing is speed: the Liouvillian
gap $\Delta_{\rm sq}\propto(1-R^2)$ closes as $R\to1$, so deeper targets converge
more slowly and require longer coherence to reach their floor. 

For the quadratic jump the distribution splits off the pole into two lobes with
interference fringes between them [Fig.~\ref{fig:ideal}(a)], and the fidelity rises to a high plateau over
a time $\sim1/\Delta_{\rm cat}$ [Fig.~\ref{fig:ideal}(c)]. The drive $\Omega$
sets the size $|\beta|^2$, and the protocol stabilizes both sizes
shown, $|\beta|^2\approx5.5$ and $\approx11$, within $10^4$ cycles. The fully polarized state has even parity, and hence so does the stabilized cat. 

Unlike the squeezed state, the cat manifold is degenerate: $I_-^2-\alpha^2$
annihilates the even and the odd cat alike. The gap $\Delta_{\rm cat}$ protects
this manifold from outside but not within it. The cat parity is set by the
initial state, which is even for the polarised pole. Residual
higher-order Floquet terms, odd in $I_\pm$, couple the two parities, and with no
gap to suppress them they drive a slow drift [Fig.~\ref{fig:ideal}(c),
$\langle P\rangle$]. It sets in only beyond $\sim1$s ($10^7$ cycles) and stays below $10^{-3}$
over the full simulated range of $\sim20$s ($10^8$ cycles), so the even cat is
the steady state on any experimental timescale.

For both squeezed and cat states the full stroboscopic dynamics match the effective master equations
of Eqs.~\eqref{eq:Lsq} and~\eqref{eq:cat_me}, plotted as the
solid curves in Fig.~\ref{fig:ideal}(b,c). This confirms numerically that our stroboscopic construction matches the expected dissipative dynamics on the nuclear spin ensemble with the electron adiabatically eliminated.\\
\begin{figure}
  \centering
  \includegraphics[width=\linewidth]{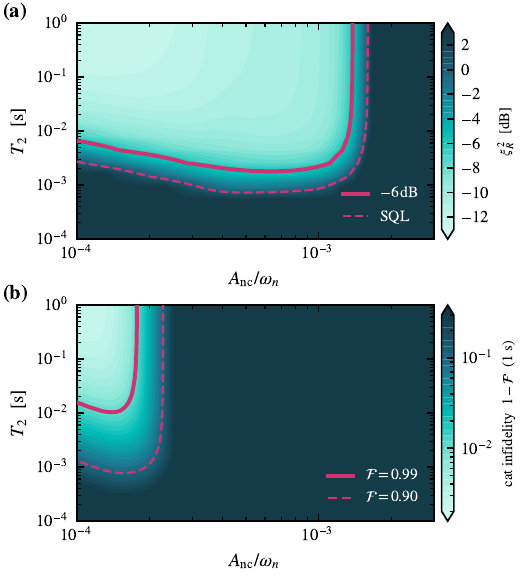}
  \caption{\textbf{Robustness to nuclear dephasing.} Quality parameter at $1$s
  under collective dephasing over the coupling-coherence plane, from the full
  stroboscopic simulation. Each region is bounded on the left where the engineered Liouvillian gap drops below the dephasing rate. Stronger coupling widens the gap and tolerates shorter $T_2$, up to a strong-coupling edge where the Floquet engineering breaks down. 
  \textbf{(a)} Squeezing $\xi^2_R$ [dB] at $R=0.90$; solid/dashed mark $-6$dB and the SQL. The floor is $-12.6$dB. 
  \textbf{(b)} Cat infidelity $1-\mathcal{F}$ at fixed size $|\beta|^2\approx5.5$ (colour,
  log scale); solid and dashed lines mark $\mathcal{F}=0.99$ and $\mathcal{F}=0.90$ in a narrow high-fidelity window around $A_{\rm nc}/\omega_n\sim1.5\times10^{-4}$. Throughout, $\omega_n/2\pi=100$MHz, $A_{\rm c}/\omega_n=0.0011$ and $I=10^4$; the per-cycle measurement is set at the stabilization optimum (App.~\ref{app:HP_Bogo}), its rate scaling with the coupling and capped at $\epsilon\le0.1$.}
  \label{fig:noise}
\end{figure}
{\it Robustness to imperfections ---} Both engineered states are dissipatively protected: nuclear dephasing is
suppressed as long as its rate stays below the Liouvillian gap. Because the gap is set by the engineered
coupling, we characterize each protocol by mapping its quality parameter
($\xi^2_R$ for the squeezed state, fidelity $\mathcal F$ for the cat) at 1 second ($\sim 10^7$ cycles)
over the coupling-coherence plane $(A_{\rm nc}/\omega_n,T_2)$
(Fig.~\ref{fig:noise}). Each map is an operating region bounded by two edges: a
weak-coupling edge, where the dephasing rate exceeds the gap, and a
strong-coupling edge, where the Floquet engineering breaks down and degrades the
target even without noise.

The cat's window sits at weaker coupling than the squeezed state's, because its
interaction is a first-order Floquet term and breaks down sooner as the coupling
grows. This weaker coupling gives a smaller gap and, on its own, less
protection. It is offset by the structure of the noise. Collective dephasing
$D[I_z]$ conserves the parity $e^{i\pi I_z}$, so it cannot drive the logical
even-odd flip, and the residual error is exponentially suppressed in cat
size~\cite{Mirrahimi.2014,Guillaud.2023}, the same bias that protects cat qubits
against bit-flips. The cat therefore tolerates dephasing despite the weaker
coupling its higher Floquet order requires.

Our Floquet-engineered interactions can be straightforwardly co-designed with nuclear dynamical decoupling by interleaving electron rotations with $I_z$ inversions (App.~\ref{app:refocusing}).
A quantum-dot nuclear coherence of $T_2\approx1$ms, reachable with a single refocusing pulse~\cite{Dyte.2025},
already places both states well inside their high-quality windows (Fig.~\ref{fig:noise}): the ensemble is squeezed
to $\xi^2_R\approx-3$dB, dropping below $-6$dB by $T_2\approx1.5$ms, and the
cat reaches $\mathcal F\approx0.95$. With more dynamical decoupling pulses that suppress residual quadrupolar and dipolar broadening, coherence times in GaAs quantum dots can reach beyond $100$ms~\cite{Dyte.2025}. At $T_2\approx100$ms the
squeezing reaches its floor, approaching the analytic
$(1-R)/(1+R)\approx-12.8$dB, and the cat attains its high-fidelity window.

Two forms of ensemble disorder leave the stabilization intact
(App.~\ref{app:disorder}). First, non-uniform hyperfine couplings do not change the
mechanism: the jump operators are simply built from a disorder-weighted
collective mode ~\cite{Villazon.20208p} that still closes as a single bright mode, at the cost of a small
imperfection $\eta\simeq n_b/(I_0 N)$ set by the excitation density $n_b$ (the
mean number of spin flips in the target state), which is kept negligible by working near the poles $\lvert I,\pm I\rangle$. Second, frequency disorder, modelled as independent local dephasing $\sqrt{\gamma}\sigma_z^{(k)}$ on each spin, drives $I$-drift: population slowly migrates from the $I_{\rm max}=N/2$ sector into lower-$I$ sectors. Crucially, the engineered dissipation continues to act within each sector it reaches, maintaining the intra-sector dark state throughout. The quality metric nonetheless degrades over time, because the mean spin $\langle \vec I \rangle $ falls below its initial value $N/2$ while the standard quantum limit reference is fixed to the total spin $N$. The degradation timescale is $\tau^*\sim N/\gamma$, since reaching the low-$I$ sectors requires $O(N)$ successive steps down the $I$-ladder; larger ensembles are therefore more robust. With $\gamma=2/T_2$ and the register of Ref.~\cite{Appel.2025} ($N\sim2\times10^4$), this is of order $10$\,s at $T_2\approx1$\,ms and $\sim10^3$\,s once dynamical decoupling reaches $T_2\approx100$\,ms, placing the frequency-disorder degradation time well beyond the formation and stabilisation times demonstrated here (App.~\ref{app:disorder}).

{\it Conclusions---} We have introduced a stroboscopic protocol that makes Floquet Hamiltonian
engineering compatible with reservoir engineering by separating the two in
time: a coherent control window shapes an effective interaction, and a
subsequent weak measurement of an auxiliary spin imprints a dissipative
channel, with the engineered Hamiltonian and jump operator set independently by
the two halves of each cycle. In a central-spin realization, adiabatic
elimination of the weakly measured electron leaves an effective master equation
for the nuclear ensemble whose collective jump operator is programmed by the
electron pulse sequence. The same construction yields two qualitatively different
many-body states, set by the collective jump operator that is engineered: a
single-quantum exchange stabilizes a Gaussian spin-squeezed state, while a
two-quantum exchange synthesized as the leading
correction of the high-frequency expansion stabilizes a non-Gaussian
collective cat. Full stroboscopic simulations confirm both, and show that each
target is protected by a finite Liouvillian gap that renders it robust to
collective and local dephasing once the engineered rate exceeds the error rate.

Crucially, this condition is met in current semiconductor quantum dots. The scheme uses only established primitives of these systems: pulse-sequence control of the electron-nuclear exchange and weak optical measurement of the electron~\cite{Gangloff.2019,Appel.2025}. Its realization therefore requires no new capability. With nuclear dynamical decoupling, co-designed and compatible with our construction, $T_2$ can reach $100$\,ms and the dissipation outpaces the dephasing by more than an order of magnitude: the states form in a few milliseconds and are held for a full second of storage, the squeezing approaching its analytic floor of $-12.8$dB and the cats, of size up to $|\beta|^2\approx 11$, exceed $0.99$ fidelity. The two states are resources for complementary tasks. The spin-squeezed state offers a
metrological gain, its sub-SQL fluctuations enhancing phase estimation; the cat, a
parity-protected even/odd code space stabilized autonomously in a collective nuclear
spin, is a candidate both for a protected quantum memory and, as a hardware-efficient
cat qubit, for quantum error correction~\cite{Franke.2026}.

Because our construction requires neither the target interaction nor the
dissipative channel to be natively available, it extends naturally to other
ancilla-controlled collective systems~\cite{Greiner2017,Ruskuc2021a} and to higher-order jump operators
engineered at higher Floquet orders, opening a route to dissipatively
stabilizing a broader family of nonclassical many-body states.

{\it Acknowledgements---}
We thank Christian Schimpf and Yusuf Karli for critical reading of our manuscript. C.F. acknowledges funding from the German Academic Scholarship Foundation (Studienstiftung des deutschen Volkes), the Royal Society, and a Cavendish Laboratory scholarship. D.G. acknowledges a Royal Society University Research Fellowship and the QuantERA project MEEDGARD through EPSRC EP/Z000556/1. We acknowledge support from the University of Cambridge's High Performance Computing facilities.

\bibliography{Dissipation_Paper}

\appendix
\section{Stroboscopic implementation of Lindblad dynamics}
\label{app:framework}

In this appendix we derive the stroboscopic protocol used in the main
text, showing how alternating coherent evolution and weak measurement
on an auxiliary spin reproduces a target Lindblad evolution in the
coarse-grained limit. The construction is a repeated-interaction (collision) model~\cite{Ciccarello.2022} with a measured auxiliary; we collect the standard steps here to fix notation and specialize them to the central-spin setting. Stroboscopic ancilla-based protocols have been used to steer many-body systems into states with frustration-free parent Hamiltonians, where local mapping operators can be written down directly and the ancilla is projectively reset each cycle~\cite{Roy.2020,Puente.2024}. Our construction differs in both ingredients: the jump operator is not given but synthesized, by Floquet-engineering an ancilla-target interaction that is not natively available, and the ancilla is only weakly measured, so the coarse-grained dynamics are Lindbladian by design rather than a limit to be escaped.

Each cycle has duration $\tau$ and consists of two steps. In the
first, the system evolves unitarily under an engineered Hamiltonian
$H$,
\begin{equation}
    \rho \mapsto  e^{-iH\tau}\rho e^{iH\tau}
    = \rho - i\tau[H,\rho] + \mathcal{O}(\tau^2).
    \label{eq:coherent_step}
\end{equation}
In the second, a weak measurement is performed on the auxiliary,
described by the two Kraus operators \cite{Wiseman.2009}
\begin{align}
    K_0 &= \mathbb{I} - \tfrac{\epsilon}{2}L^\dagger L
           + \mathcal{O}(\epsilon^2),
    \label{eq:K0}
    \\
    K_1 &= \sqrt{\epsilon}L,
    \label{eq:K1}
\end{align}
with $\epsilon\ll1$ quantifying the measurement strength. The form of
$K_0$ is fixed by the completeness relation
$K_0^\dagger K_0 + K_1^\dagger K_1 = \mathbb{I}$, expanded to first order
in $\epsilon$. 

The operator $L$ is fixed by the measurement: in the central-spin realization it is the electron lowering operator, $L=S_-$. The weak measurement is a short pulse of spin-selective optical pumping such that with a small probability $\epsilon$ per cycle a photon is scattered and the electron relaxes to $\ket{\downarrow}$, while $\ket{\downarrow}$ is left untouched. Discarding the photon record, this gives the Kraus operators $K_0=\sqrt{1-\epsilon}\ket{\uparrow}\bra{\uparrow}+\ket{\downarrow}\bra{\downarrow}$ and $K_1=\sqrt{\epsilon}\ket{\downarrow}\bra{\uparrow}$, these reduce to Eqs. \ref{eq:K0} and \ref{eq:K1} with $L=S_-$.

Tracing out the measurement record yields the unconditional update
\begin{equation}
    \rho' = K_0\rho K_0^\dagger + K_1\rho K_1^\dagger.
\end{equation}
Substituting Eqs.~\eqref{eq:K0}--\eqref{eq:K1} and keeping terms to
leading order in $\epsilon$ gives
\begin{equation}
    \rho' = \rho + \epsilon\mathcal{D}[L]\rho
              + \mathcal{O}(\epsilon^2),
    \qquad
    \mathcal{D}[L]\rho = L\rho L^\dagger - \tfrac{1}{2}\{L^\dagger L,\rho\}.
    \label{eq:meas_step}
\end{equation}
Combining the coherent step \eqref{eq:coherent_step} with the
measurement step \eqref{eq:meas_step}, the density matrix at the end of
one cycle is
\begin{equation}
    \rho(t+\tau)
    =
    \rho(t)
    - i\tau[H,\rho(t)]
    + \epsilon\mathcal{D}[L]\rho(t)
    + \mathcal{O}(\tau^2,\epsilon^2,\epsilon\tau).
    \label{eq:single_cycle}
\end{equation}

Equation~\eqref{eq:single_cycle} is a finite-difference update. To
recover a continuous master equation we take the limit
\begin{equation}
    \epsilon,\tau \to 0,
    \qquad
    \Gamma \equiv \epsilon/\tau \text{ fixed},
\end{equation}
which corresponds to many weak cycles per characteristic timescale of
the engineered dynamics. Dividing Eq.~\eqref{eq:single_cycle} by
$\tau$ and dropping the subleading terms yields the Lindblad master
equation
\begin{equation}
    \dot\rho = -i[H,\rho] + \Gamma\mathcal{D}[L]\rho,
    \label{eq:lindblad_app}
\end{equation}
which is Eq.~\eqref{eq:lindblad} of the main text. The Hamiltonian $H$
is set by the coherent-control window, while the dissipation rate
$\Gamma$ and jump operator $L$ are set by the weak measurement; the
two are engineered independently within the cycle.

The expansion underlying Eq.~\eqref{eq:single_cycle} assumes that the
state changes only weakly per cycle~\cite{Gamel.2010}. Concretely, the neglected terms
are small when
\begin{equation}
    \|H\|\tau \ll 1,
    \qquad
    \epsilon \ll 1,
    \qquad
    \|H\|\tau \cdot \epsilon \ll \max(\|H\|\tau,\epsilon),
    \label{eq:validity}
\end{equation}
i.e.\ when neither the unitary kick nor the measurement back-action is
strong enough on its own to take the state outside the linear regime
within a single cycle. In our central-spin implementation, $\|H\|\tau$
is controlled by the ratio $A_{\rm nc}/\omega_n$ of the transverse
hyperfine coupling to the nuclear Larmor frequency, while $\epsilon$
is controlled by the readout integration time and probe coupling. Both are tuneable and
can be made small simultaneously, as confirmed numerically in
the main text, where stroboscopic and continuous
trajectories collapse onto each other as $A_{\rm nc}/\omega_n \to 0$.

So far $H$ and $L$ act on the joint auxiliary-target space. When the measurement
damps the auxiliary much faster than the engineered coupling drives it, the auxiliary
can be adiabatically eliminated, and standard projection-operator
techniques~\cite{Reiter.2012} give an effective master equation for the target
state $\rho_n$ alone,
\begin{equation}
    \dot\rho_n = -i[H_{\rm eff},\rho_n] + \Gamma_{\rm eff}
                     \mathcal{D}[L_{\rm eff}]\rho_n,
    \label{eq:eff}
\end{equation}
with $H_{\rm eff}$ and $L_{\rm eff}$ set by the engineered interaction. Here the
auxiliary is the electron, damped at its measurement rate $\kappa$; the elimination
produces the collective nuclear jump operators
$L_{\rm sq}\propto\Omega_+ I_- + \Omega_- I_+$ and $L_{\rm cat}\propto I_-^2-\alpha^2$
of the squeezing and cat protocols.

The fixed points of Eq.~\eqref{eq:eff} satisfy
$\dot\rho_{ss}=0$. When $H$ leaves the kernel of $L_{\rm eff}$ invariant, any
state in $\ker L_{\rm eff}$ is a steady state; the dynamics within this kernel
are then unitary and selected by $H$. If, in addition, the combined
action of $H$ and $\mathcal{D}[L_{\rm eff}]$ leaves no nontrivial subspace
invariant (i.e.\ the dynamics are irreducible), the steady state is
unique~\cite{Frigerio.1978}. Both target states constructed in the
main text, the spin-squeezed state and
the cat state, arise as the kernel of an
engineered $L$ in this way. For the squeezed state $\ker L_{\rm eff}$ is
one-dimensional and the steady state is unique. The two-photon cat dissipator $L_{\rm cat}\propto I_-^2-\alpha^2$ is different:
it conserves the spin-flip parity $\Pi=(-1)^{I-I_z}$, so its kernel is
two-dimensional, spanned by the even and odd cats, and the two parity sectors do not mix. The steady state is therefore not unique in general, but within each parity sector it is: because $\Pi$ is conserved, the dynamics remain in the sector fixed by the initial state. Initializing at the fully polarized pole $|I,-I\rangle$, which has definite parity, the protocol converges to the even cat, to which we quote fidelities throughout. 

\section{Floquet engineering of the electron-nuclear coupling}
\label{app:eng}

The protocol separates in time the two ingredients of the effective dissipative
dynamics: a periodic electron pulse sequence shapes a coherent electron-nuclear
coupling (this appendix), and a weak measurement of the electron converts it into
a collective nuclear jump operator (Appendix~\ref{app:framework}). The pulse sequence
fixes the order of the nuclear operator that appears in the coupling. A
single-quantum exchange survives already at zeroth order in the high-frequency
expansion and stabilizes a squeezed state; a two-quantum coupling is
absent at zeroth order and is instead generated at first order, and
stabilizes a cat. We first set up the general framework with the full
hyperfine interaction and all three control axes, then specialise to each case.

\subsection{General framework}
\label{app:eng_fw}

A time-periodic Hamiltonian $H(t+T)=H(t)$ with Fourier components
$H(t)=\sum_n H^{(n)}e^{in\omega t}$, $\omega=2\pi/T$, is captured at long times by
the van~Vleck effective Hamiltonian~\cite{Rahav.2003,Goldman.2014}, a power series in $1/\omega$,
\begin{align}
  H_F^{(0)} &= H^{(0)},
  \label{eq:vv0}\\
  H_F^{(1)} &= \sum_{n\geq1}\frac{1}{n\omega}[H^{(n)},H^{(-n)}],
  \label{eq:vv1}
\end{align}
with higher orders suppressed by further powers of $1/\omega$. It needs no time
ordering and is well suited to piecewise-constant pulse sequences.

We apply it to a central electron spin $\mathbf S$ coupled to a collective
nuclear spin $\mathbf I$ through the full hyperfine interaction,
\begin{equation}
  H(t)=\omega_n I_z + A_cS_z I_z + A_{\rm nc}S_z I_x + \Omega(t)S_x ,
  \label{eq:model}
\end{equation}
with nuclear Zeeman frequency $\omega_n$, collinear (secular) hyperfine coupling
$A_c$, non-collinear (transverse) coupling $A_{\rm nc}$, and a slow electron drive
$\Omega(t)$ used only for the cat. A train of fast electron $\pi$-pulses rotates
the coupling axis; in its toggling frame the static $S_z$ becomes a time-dependent
unit vector,
\begin{equation}
  S_z \longrightarrow \mathbf h(t)\!\cdot\!\mathbf S
  = h_x(t)S_x+h_y(t)S_y+h_z(t)S_z ,
  \label{eq:toggle}
\end{equation}
with $\mathbf h(t)$ piecewise constant and pointing along one Cartesian axis per
slot. All three components $h_{x,y,z}$ enter the general construction; squeezing
will need only $h_x,h_y$, while the cat requires $h_z$ as well.

Moving to the nuclear rotating frame $U_0=e^{-i\omega_n t I_z}$, the collinear
term is static ($I_z$ commutes with $U_0$) while the transverse term acquires the
nuclear phase, $I_x\to\tfrac12(I_+e^{i\omega_n t}+I_-e^{-i\omega_n t})$:
\begin{equation}
  \tilde H(t)=\bigl[\mathbf h(t)\!\cdot\!\mathbf S\bigr]
  \Bigl[A_c I_z+\tfrac{A_{\rm nc}}{2}\bigl(I_+e^{i\omega_n t}
  +I_-e^{-i\omega_n t}\bigr)\Bigr].
  \label{eq:Hrot}
\end{equation}
We expand each control in a Fourier series and collect the three coefficients
into one complex vector,
\begin{equation}
  h_a(t)=\sum_\ell (\mathbf C_\ell)_ae^{i\ell\omega t},\quad
  \mathbf C_\ell=(X_\ell,Y_\ell,Z_\ell),\quad \omega=\frac{\omega_n}{r},
  \label{eq:fourier}
\end{equation}
with $r\in\mathbb Q$ and, for real controls, $\mathbf C_{-\ell}=\mathbf C_\ell^{*}$.
The hyperfine phase shifts the harmonic index of the transverse term by $\mp r$,
while the collinear term keeps its bare index, so $\tilde H(t)=\sum_\ell
H^{(\ell)}e^{i\ell\omega t}$ with
\begin{equation}
  H^{(\ell)}=A_c(\mathbf C_{\ell}\!\cdot\!\mathbf S)I_z
  +\frac{A_{\rm nc}}{2}\Bigl[(\mathbf C_{\ell-r}\!\cdot\!\mathbf S)I_+
  +(\mathbf C_{\ell+r}\!\cdot\!\mathbf S)I_-\Bigr].
  \label{eq:Hl}
\end{equation}
A term contributes to $H_F^{(0)}$ only if its total phase is static: at $\ell=0$
for the collinear $I_z$ piece, and at $\ell=\pm r$ for the single-quantum $I_\pm$.
We use a zero-mean sequence, $\mathbf C_0=0$, which removes the static collinear
contribution. A rigid shift of one control relative to another by a fraction
$\nu$ of the period imprints a phase,
\begin{equation}
  h_y(t)=h_x(t+\nu T)\Longrightarrow Y_\ell=e^{i2\pi\nu\ell}X_\ell ,
  \label{eq:shift}
\end{equation}
the lever used below to select the handedness of the exchange. The concrete
sequences are shown in Fig.~\ref{fig:app_seq}.

\subsection{Squeezing: single-quantum exchange at zeroth order}
\label{app:eng_sq}

With $\mathbf C_0=0$, the zeroth-order Floquet Hamiltonian is the resonant
($\ell=\pm r$) single-quantum exchange,
\begin{equation}
  H_F^{(0)}=\frac{A_{\rm nc}}{2}\Bigl[(\mathbf C_{-r}\!\cdot\!\mathbf S)I_+
  +(\mathbf C_{r}\!\cdot\!\mathbf S)I_-\Bigr],
  \label{eq:H0sq}
\end{equation}
and for squeezing it suffices to keep the control in the $x$--$y$ plane
($Z_\ell=0$). Two eight-slot primitive sequences (Fig.~\ref{fig:app_seq}, lower
panels) share the same $h_y$ and differ only by a shift of $h_x$:
\begin{align}
  \text{flip-flop:}\quad
  h_x&=(0,0,\text{-}1,1,0,0,1,\text{-}1),\nonumber\\
  h_y&=(\text{-}1,1,0,0,1,\text{-}1,0,0),
  \label{eq:ffseq}\\[2pt]
  \text{flip-flip:}\quad
  h_x&=(0,0,1,\text{-}1,0,0,\text{-}1,1),
  \label{eq:flseq}
\end{align}
with $h_y$ common. The flip-flop has $h_y(t)=h_x(t+T/4)$, so by
Eq.~\eqref{eq:shift} $Y_\ell=i^\ell X_\ell$; at $\ell=r$ this gives
$X_r-iY_r=0$, which selects the cross-handed exchange. The flip-flip has
$h_y(t)=h_x(t+3T/4)$, so $Y_\ell=(-i)^\ell X_\ell$ and $X_r+iY_r=0$, selecting the
same-handed exchange:
\begin{align}
  H^{(0)}_{\uparrow\downarrow}&\propto S_+I_-+S_-I_+
  &&\text{(flip-flop)},\nonumber\\
  H^{(0)}_{\uparrow\uparrow}&\propto S_+I_++S_-I_-
  &&\text{(flip-flip)}.
  \label{eq:fffl}
\end{align}

\paragraph{Tuning the anisotropy by concatenation.}
Each primitive is a complete Floquet period with a single, definite zeroth-order Hamiltonian, Eq.~\eqref{eq:fffl}: we never mix the two interactions \emph{within} a period. A macro-cycle is assembled from $N_{\uparrow\uparrow}$ flip-flip and $N_{\uparrow\downarrow}$ flip-flop periods, with the target ratio $R = N_{\uparrow\uparrow}/N_{\uparrow\downarrow}$ realized by a rational approximation and the two period types interleaved as uniformly as possible across the macro-cycle (an error-diffusion ordering that spreads the flip-flip periods evenly among the flip-flop periods), rather than run as two contiguous blocks. Both period types share the same fundamental period $T$, so over the macro-cycle each contributes its own constant $H^{(0)}$ for the fraction of time it occupies; the uniform interleaving keeps the excursions between the two interactions short, minimizing the higher-order corrections to the $R$-weighted average that sets the engineered interaction. The leading
(zeroth-order Magnus~\cite{Blanes.2009}) average over the macro-cycle is therefore the time-weighted
sum,
\begin{equation}
  H_{\rm eff}=\frac{N_{\uparrow\uparrow}H^{(0)}_{\uparrow\uparrow}
  +N_{\uparrow\downarrow}H^{(0)}_{\uparrow\downarrow}}
  {N_{\uparrow\uparrow}+N_{\uparrow\downarrow}},
  \label{eq:macro}
\end{equation}
which gives the target
\begin{equation}
  H_{\rm eff}=\Omega_-(S_+I_++S_-I_-)+\Omega_+(S_+I_-+S_-I_+),
  \label{eq:Hsq}
\end{equation}
with the ratio set purely by an integer count of periods,
\begin{equation}
  \frac{\Omega_-}{\Omega_+}=\frac{N_{\uparrow\uparrow}}{N_{\uparrow\downarrow}} .
  \label{eq:ratio}
\end{equation}
The engineered anisotropy, and hence the squeezing strength, is thus dialled by
how many periods of each primitive are run, while the intra-period $\ell=\pm r$
selection is left untouched. A weak measurement of the electron and adiabatic
elimination (Appendix~\ref{app:framework}) turn Eq.~\eqref{eq:Hsq} into the collective
single-quantum jump $L_{\rm sq}\propto\Omega_+I_-+\Omega_-I_+$, whose dark state
is a spin-squeezed (Bogoliubov-vacuum) state.

\subsection{Cat: two-quantum coupling at first order}
\label{app:eng_cat}

For a cat the single-quantum exchange must be absent, so the sequence carries no
resonant ($\ell=\pm r$) harmonic and the leading coupling is the first-order term
$H_F^{(1)}$ of Eq.~\eqref{eq:vv1}. We write this commutator in full and then go
through it term by term, showing how one sequence keeps the wanted piece and removes
the rest.

\paragraph{The first-order commutator.}
Inserting Eq.~\eqref{eq:Hl} into Eq.~\eqref{eq:vv1} and grouping the per-harmonic
commutator $[H^{(\ell)},H^{(-\ell)}]=T^{(2)}_\ell+T^{(1)}_\ell+T^{(0)}_\ell$ by the
nuclear change $\Delta m$, each coefficient is a bilinear in the Fourier vectors
$\mathbf C_n=(X_n,Y_n,Z_n)$ (with $\mathbf C_{-n}=\mathbf C_n^{*}$):
\begin{align}
  T^{(2)}_\ell &= i A_{\rm nc}^{2}
     (\mathbf C_{\ell-r}\!\times\!\mathbf C_{\ell+r}^{*})\!\cdot\!\mathbf SI_+^2
     +\text{h.c.},
     \label{eq:T2}\\[2pt]
  T^{(0)}_\ell &= i\bigl[A_{\rm nc}^{2}(\mathbf C_{\ell-r}\!\times\!\mathbf C_{\ell-r}^{*}
        +\mathbf C_{\ell+r}\!\times\!\mathbf C_{\ell+r}^{*})
        +A_c^{2}\mathbf C_{\ell}\!\times\!\mathbf C_{\ell}^{*}\nonumber\\&\quad\bigr]\!\cdot\!\mathbf SI_z^2
      +\tfrac12 A_{\rm nc}^{2}\bigl(|\mathbf C_{\ell+r}|^{2}-|\mathbf C_{\ell-r}|^{2}\bigr)I_z,
     \label{eq:T0}\\[2pt]
  T^{(1)}_\ell &= \tfrac{i}{2}A_cA_{\rm nc}\bigl(\mathbf C_{\ell-r}\!\times\!\mathbf C_{\ell}^{*}
        +\mathbf C_{\ell}\!\times\!\mathbf C_{\ell+r}^{*}\bigr)\!\cdot\!\mathbf S\{I_z,I_+\}
     \nonumber\\
     &\quad +\tfrac14 A_cA_{\rm nc}\bigl(\mathbf C_{\ell}\!\cdot\!\mathbf C_{\ell+r}^{*}
        -\mathbf C_{\ell-r}\!\cdot\!\mathbf C_{\ell}^{*}\bigr)I_+ +\text{h.c.}.
     \label{eq:T1}
\end{align}
The three
coefficients carry three distinct bilinear Fourier components: a cross product of two
harmonics separated by $2r$ (in $T^{(2)}_\ell$), a self cross product together with a
magnitude difference ($T^{(0)}_\ell$), and cross and scalar products of harmonics
separated by $r$ ($T^{(1)}_\ell$). Because of this difference in structure we can design a sequence that keeps $T^{(2)}_\ell$
while cancelling the rest.

\begin{figure}
  \centering
  \includegraphics[width=\linewidth]{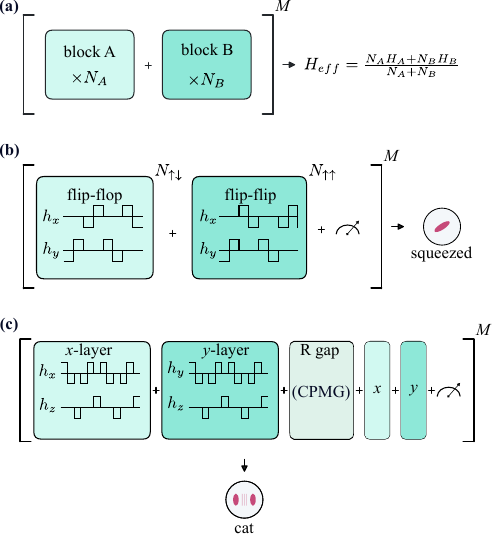}
  \caption{\textbf{Pulse sequences} (toggling-frame controls $\mathbf h(t)$).
  \textbf{(a)} General principle: a macro-cycle concatenates $N_A$ periods of block~A
  and $N_B$ of block~B and repeats $\ell$ times; to leading (zeroth-order Magnus) order the
  engineered Hamiltonian is the time-weighted average
  $H_{\rm eff}=(N_AH_A+N_BH_B)/(N_A+N_B)$ [Eq.~\eqref{eq:macro}].
  \textbf{(b)} Squeezing: two $8$-slot primitives over one period $T$, sharing a common
  $h_y$ and differing by a $T/4$ shift of $h_x$ (solid: flip-flop, $S_+I_-{+}S_-I_+$;
  dashed: flip-flip, $S_+I_+{+}S_-I_-$); interleaving $N_{\uparrow\downarrow}$ flip-flop
  and $N_{\uparrow\uparrow}$ flip-flip periods with a weak measurement stabilizes the
  squeezed state.
  \textbf{(c)} Cat: the engineered train $U_\varphi$ (period $2T$) runs an $x$-transverse
  layer ($H_2$, $t/T\in[0,1]$) then a $y$-transverse layer ($H_1$, $t/T\in[1,2]$), with
  $z$-pulses interleaved throughout ($h_z$) to join the two quadratures into the chiral
  coupling. The full cycle runs the train at clock phase $\varphi=0$ (pulses shown to
  scale), a symmetriser gap $R$ [a half-Larmor dead time $\tau=\pi/\omega_n$ carrying $8$
  electron $\pi_x$ echoes (CPMG) that refocus the hyperfine, drawn schematically], the
  train at $\varphi=\pi$ (drawn compactly as $x/y$ blocks, colour-coded as in the first
  train), and the weak electron measurement; every second gap is skipped so that one
  parity flip occurs per cycle (an odd $R$-count), refocusing the reopened single-quantum
  term.}
  \label{fig:app_seq}
\end{figure}

\paragraph{Term by term.}
The realization used in the simulations [Fig.~\ref{fig:app_seq}(b)] makes this
concrete. It is a period-$2T$ train of two layers: an $x$-transverse layer and a
$y$-transverse layer, each of duration $T$ and split into four quarters. Within a
quarter the transverse pulse pair is $(+,-)$ and the quarter sign alternates, giving
the transverse pattern $+,-,-,+$, while the two $z$-pulses, of slightly unequal
duration, repeat identically every quarter. The transverse control is then
anti-invariant and the longitudinal control invariant under a quarter shift,
\begin{equation}
  h_{x,y}\!\left(t+\tfrac{T}{4}\right)=-h_{x,y}(t),\qquad
  h_{z}\!\left(t+\tfrac{T}{4}\right)=+h_{z}(t),
  \label{eq:quartershift}
\end{equation}
so, since a harmonic acquires $i^{\ell}$ under $t\to t+T/4$, only certain harmonics
survive: the transverse controls are pinned to $\ell\equiv2\pmod4$ and the $z$ controls to
$\ell\equiv0\pmod4$. 

\emph{Two-photon} ($T^{(2)}_\ell$, the target). The cross-harmonic product
$\mathbf C_{\ell-r}\!\times\!\mathbf C_{\ell+r}^{*}$ needs two harmonics separated by
$2r$, hence at least two pulse axes, and lies in the $x$--$y$ plane only with
nonzero $h_z$. Pairing a transverse harmonic ($\ell\equiv2$) with a $z$ harmonic
($\ell\equiv0$), which differ by $2r\equiv2\pmod4$ for $r=5$, both factors are
allowed, so the product survives.

The target coupling $S_+I_-^2+S_-I_+^2$ is \emph{chiral}: the electron raising
operator $S_+$ is tied to one sense of the nuclear two-quantum process ($I_-^2$) and
$S_-$ to the other ($I_+^2$). It splits into two Cartesian pieces,
\begin{equation}
  S_+I_-^2+S_-I_+^2
  = \underbrace{S_x\!\left(I_+^2+I_-^2\right)}_{H_1}
  - \underbrace{iS_y\!\left(I_+^2-I_-^2\right)}_{H_2},
  \label{eq:H1H2}
\end{equation}
each of which is individually \emph{non-chiral}, so that only their sum with the relative factor $i$, selects a definite
handedness. A single three-axis sequence cannot enforce the selection rules for both
pieces at once, so we engineer $H_1$ and $H_2$ in separate layers, the $y$- and
$x$-transverse layers respectively, and combine them through the outer Magnus
expansion, exactly as for squeezing. We take $r=5$, the harmonic with the strongest
coefficients.

The relative phase that renders the combination $H_1-H_2$ chiral is
supplied by a rigid offset $\nu$ of the $x$-layer relative to the
$y$-layer. By Eq.~\eqref{eq:shift} the offset multiplies the
$x$-layer harmonics by $e^{i2\pi\nu\ell}$; since the two-photon term
pairs harmonics at $\ell\mp r$ [Eq.~\eqref{eq:T2}], separated by $2r$,
its coefficient acquires the net phase
\begin{equation}
  \mathbf{C}_{\ell-r}\times\mathbf{C}^{*}_{\ell+r}
  \longrightarrow
  e^{i4\pi r\nu}
  \mathbf{C}_{\ell-r}\times\mathbf{C}^{*}_{\ell+r},
  \label{eq:offset-phase}
\end{equation}
which rotates $H_2$ relative to the fixed $H_1$. Writing the deviation
of the coupling from the chiral direction $S_-$ as
$V_y/V_x=-ie^{i\delta}$, the chirality phase is linear in the offset,
\begin{equation}
  \delta(\nu)=\delta_0+4\pi r\nu,
  \qquad
  \nu^{\ast}=-\frac{\delta_0}{4\pi r}=\frac{1}{8r},
  \label{eq:chirality-law}
\end{equation}
where $\delta_0=-\pi/2$ and the chiral point $\delta=0$ sits at
$\nu^{\ast}=1/(8r)=0.025$ for $r=5$. At $\nu=0$ the two layers are in
phase, $\mathbf{V}$ is real, and the coupling is non-chiral, with equal
weight on $I_+^2$ and $I_-^2$, so no cat forms; the sign of the offset
selects the handedness, $\nu\to-\nu$ exchanging $I_-^2\leftrightarrow
I_+^2$. Equation~\eqref{eq:chirality-law} is the two-quantum analogue
of the squeezing's $T/4$ flip-flop shift, finer by a factor $2r$
because the two-photon term feels the harmonic separation $2r$ rather
than the single-quantum's $\pm r$.

\emph{Diagonal self-products} ($T^{(0)}_\ell$). Because each layer uses a single
transverse axis and the two harmonic classes sit at different $\ell$, every harmonic
is uniaxial, purely transverse or purely $z$, so the self cross products
$\mathbf C_n\!\times\!\mathbf C_n^{*}$ vanish identically.

\emph{Detuning} ($T^{(0)}_\ell$). The one residual term these selection rules leave
is the $I_z$ piece of Eq.~\eqref{eq:T0},
$\propto |\mathbf C_{\ell+r}|^2-|\mathbf C_{\ell-r}|^2$. The two sidebands
$\ell\pm r$ differ by $2r\equiv2\pmod4$, exactly the offset between the transverse
($\ell\equiv2$) and $z$ ($\ell\equiv0$) harmonics, so one sideband is carried by the
transverse pulses and the other by the $z$ pulses. Their weights therefore need not
be equal, and any mismatch leaves a spurious longitudinal ($I_z$) shift, a residual
detuning of the nuclei. Because the two weights are set by the transverse and $z$
pulse durations, equalizing them is a matter of the duration ratio: we set
$\tau_z/\tau_x\simeq0.93$, which lengthens the $z$ intervals just enough to enforce
$|\mathbf C_{\ell+r}|=|\mathbf C_{\ell-r}|$ while preserving the nuclear-Zeeman
commensurability and the quarter-shift symmetry that protect the other terms.

\emph{Single-quantum} ($T^{(1)}_\ell$, within a layer). Each contribution in
Eq.~\eqref{eq:T1} pairs harmonics separated by the \emph{odd} offset $r$: the cross
product $\mathbf C_{\ell-r}\!\times\!\mathbf C_\ell^{*}$ (and the scalar
$\mathbf C_{\ell-r}\!\cdot\!\mathbf C_\ell^{*}$) needs both $\mathbf C_{\ell-r}$ and
$\mathbf C_\ell$ nonzero, but with $r$ odd these sit at opposite parity mod $4$, one
transverse ($\ell\equiv2$), one $z$ ($\ell\equiv0$), so one factor always vanishes.
Term by term $T^{(1)}_\ell=0$: a single layer carries no first-order single-quantum
coupling.

\emph{Reopening by the chirality.} Because the target is chiral, the
train is two layers rather than one [Eq.~\eqref{eq:H1H2}], so its true
period is $2T$: the toggling fundamental halves and the harmonic that
meets the Larmor resonance is the $2r$-th, an even harmonic, where it was odd for a
single layer. At an even harmonic a hyperfine and a collinear term can
combine into a stationary $\Delta m=\pm1$ coupling, inducing a small
linear single-quantum leak $T^{(1)}\propto A_cA_{\rm nc}$.

\emph{Removal by an odd symmetriser.} The leak is parity-odd, so
the symmetriser $R=e^{-i\pi I_z}$ ($I_\pm\!\to\!-I_\pm$) flips its sign:
$R$ echoes the even/odd-cat qubit, cancelling the leak only with an
\emph{odd} number of flips per cycle, while leaving the parity-even
$\chi$ untouched. Since $U_\varphi$ already carries both chirality
layers, a gap after each train gives two flips per cycle (even) and the
leak survives; skipping every second gap leaves one (odd) and averages
it away. The gap costs no extra gate: a half-Larmor dead time
$\tau=\pi/\omega_n$ with a CPMG $\pi_x$ train (refocusing the hyperfine)
advances the nuclear phase by $\pi$, so one cycle is
$U_\varphi\to R\to U_\varphi\to$ (weak measurement) with a single gap
between. This refocusing cost grows as $A_{\rm nc}$ falls (the per-gap
kick $\sim\!A_{\rm nc}$, but the gap shrinks faster), one of the factors
fixing the operating point in Fig.~\ref{fig:noise}.

\paragraph{Drive, elimination, and validity.}
Summing $T^{(2)}_\ell$ over the two layers gives the chiral two-photon coupling of the
main text, to which the static drive $\Omega S_x$ of Eq.~\eqref{eq:model} (constant in
the toggling frame) adds:
\begin{equation}
  H_{\rm int}=\chi(S_+I_-^2+S_-I_+^2)+\Omega S_x .
  \label{eq:Hcat}
\end{equation}
A weak electron measurement and adiabatic elimination (Appendix~\ref{app:framework}) turn
this into the collective two-quantum jump $L_{\rm cat}\propto I_-^2-\alpha^2$ with
$\alpha^2=-\Omega/2\chi$, whose even dark state is the target cat. Since
$\chi\propto A_{\rm nc}^2/\omega_n$, the construction requires $A_{\rm nc}/\omega_n\ll1$:
the rate $\Gamma_{\rm cat}\sim\chi^2/\kappa\propto A_{\rm nc}^4$ rises steeply with
coupling, but at large $A_{\rm nc}$ the neglected higher-order terms, and the residual
terms not removed by symmetry, spoil the dark state even without dephasing. This
strong-coupling (Floquet-breakdown) wall bounds the operating region from the right in
Fig.~\ref{fig:noise}.

\subsection{Compatibility with nuclear refocusing}
\label{app:refocusing}
\begin{figure}
  \centering
  \includegraphics[width=\linewidth]{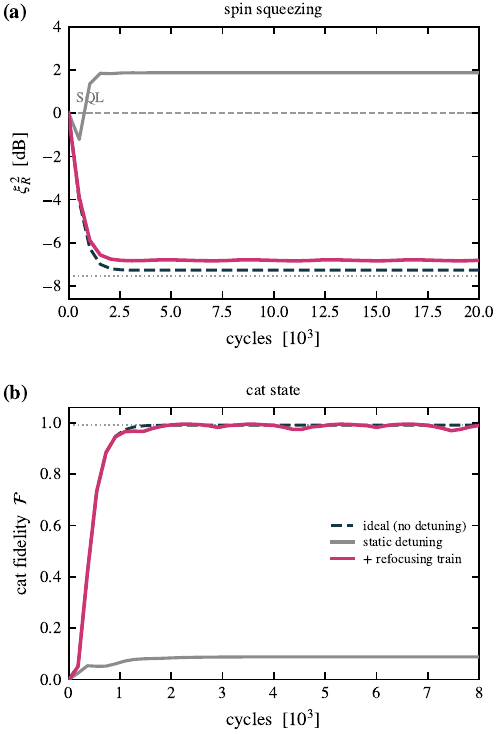}
  \caption{\textbf{A nuclear refocusing pulse train coexists with the
  stabilization.} Full stroboscopic protocol at $I = 50$ (exact $2I+1$ space, so
  the $\pi$ pulse is a genuine pole-to-pole rotation). \textbf{(a)} Spin squeezing
  $\xi_R^2(t)$, $R = 0.7$. \textbf{(b)} Cat fidelity $\mathcal{F}(t)$,
  $|\beta|^2 \approx 12$. In each panel: the ideal run with no detuning (blue
  dashed); a static collective detuning $\delta I_z$ that destroys the engineered
  state (grey); and the same detuning with a single-$\pi$ nuclear refocusing train
  interleaved, including the state-specific frame correction, a toggling-layer
  sign flip (flip-flop $\leftrightarrow$ flip-flip for squeezing; the $H_2$-layer
  flip for the cat) together with the refocusing-pulse axis of
  Eq.~\eqref{eq:phipi}, which restores the engineered state to its ideal floor
  (cherry). Time is in Floquet cycles.}
  \label{fig:refocus}
\end{figure}

The coherence time $T_2$ that sets the vertical axis of the noise map
(Fig.~\ref{fig:noise}) is the \emph{dynamically decoupled} one: in the
experiment the dominant, slowly varying nuclear dephasing is refocused by a
train of collective nuclear $\pi$ pulses, and it is the residual noise that the
stabilization must tolerate. This is only legitimate if the refocusing pulses
can be interleaved with the stabilization protocol without spoiling it. Here we
show that they can, in the simplest possible case: a single nuclear $\pi$ pulse
per refocusing interval. Any more elaborate sequence (CPMG, \dots) is a
concatenation of this elementary step.

\subsubsection{Frame accounting}

A collective nuclear $\pi$ pulse about a transverse axis inverts the ensemble,
\begin{equation}
  I_z \to -I_z, \qquad I_\pm \to I_\mp ,
\label{eq:pi_map}
\end{equation}
which is precisely what refocuses a quasi-static collective detuning
$\delta I_z$: the phase accumulated in one interval is undone in the next.
However, the same inversion also acts on the engineered jump operator. For the
squeezed state the dissipator is $\mathcal{D}[L_{\rm sq}]$ with
\begin{equation}
  L_{\rm sq} = \cos\theta I_+ + \sin\theta I_-,
\end{equation}
so after a pulse Eq.~\eqref{eq:pi_map} turns it into
$\cos\theta I_- + \sin\theta I_+$, whose dark state is the \emph{anti}-squeezed
quadrature; left uncorrected, the protocol would drive the ensemble away from the
target. For the cat the dissipator is $\mathcal{D}[L_{\rm cat}]$ with
\begin{equation}
  L_{\rm cat} \propto I_-^2 - \alpha^2 ,
\end{equation}
and the pulse sends $I_-^2 \to I_+^2$, again engineering the wrong two-quantum
process.

The fix is entirely on the electron side, because the sign of the engineered
coupling is set by the drive, not by the nuclei. In both cases it is the flip of a
single drive quadrature, seen most transparently by decomposing the engineered
electron-nuclear coupling into its in-phase ($S_x$) and quadrature ($S_y$) parts.

\emph{Squeezing.} The linear coupling is
\begin{equation}
  S_+I_- + S_-I_+ = S_x(I_++I_-) - iS_y(I_+-I_-),
\end{equation}
In the pulse sequence we simply exchange the flip-flop $\leftrightarrow$ flip-flip parts of the
interleaved half-periods; the nuclear $\pi$ pulse is applied about the squeezed-quadrature
axis.

\emph{Cat.} In the same decomposition $S_+I_-^2+S_-I_+^2=H_1-H_2$
[Eq.~\eqref{eq:H1H2}], the pulse sends $I_-^2 \leftrightarrow I_+^2$, so restoring the
target on the inverted ensemble requires $S_+I_+^2 + S_-I_-^2 = H_1 + H_2$: only the
quadrature component $H_2$ changes sign, realized by negating its toggling layer,
$h_{x/z}^\prime=-h_{x/z}$.

Flipping $H_2$ fixes the chirality, but the cat is a continuous-variable code and
carries one further label, the phase-space orientation of its lobes.  The engineered offset $\alpha^2 = -\Omega/2\chi$ is complex: with a real drive
($\arg\Omega=0$) the cat's lobes sit at orientation
\begin{equation}
  \gamma \equiv \arg\alpha^2 =\pi -\arg\chi ,
\end{equation}
fixed by the phase of the two-quantum coupling $\chi$. The refocusing $\pi$ must reproduce the
cat at this orientation on the inverted pole. Taking the pulse about a transverse
axis at in-plane angle $\varphi$, $R(\varphi)=e^{-i\pi I_\varphi}$ with
$I_\varphi = \cos\varphi I_x + \sin\varphi I_y$, and writing
$R(\varphi)=e^{-i\varphi I_z}e^{-i\pi I_x}e^{i\varphi I_z}$, one finds
\begin{equation}
  R(\varphi)I_+R(\varphi)^\dagger = e^{2i\varphi} I_- ,
  \qquad
  R(\varphi)I_+^2R(\varphi)^\dagger = e^{4i\varphi} I_-^2 ,
  \label{eq:Rphi}
\end{equation}
so the $\pi$-image of the dark cat is stabilized by
$R(I_-^2-\alpha^2)R^\dagger \propto I_+^2 - e^{4i\varphi}\alpha^2$: a top-pole cat
with effective offset $\alpha^2_{\rm img}=e^{4i\varphi}\alpha^2$. The $H_2$ flip is
a quadrature reflection, so the correction engineers the conjugate offset,
$\alpha^2_{\rm corr}=(\alpha^2)^\ast$. Matching
$\alpha^2_{\rm corr}=\alpha^2_{\rm img}$ gives $e^{4i\varphi}=e^{-2i\gamma}$, i.e.
\begin{equation}
  \varphi_\pi = -\tfrac12\gamma = \tfrac12\arg\chi
  \pmod{\tfrac{\pi}{2}}
  \label{eq:phipi}
\end{equation}
The refocusing axis is therefore half the
coupling phase. 

Both corrections thus use only physical pulse controls: an electron-toggling sign
(the chirality flip) and one RF phase on the refocusing pulse (the axis). No
manipulation of $\omega_n$ and no auxiliary collective rotation is required.

\subsubsection{Demonstration}

Figure~\ref{fig:refocus} shows the interleaved dynamics for both states. We run the full stroboscopic protocol at a reduced spin
$I = 50$ in the complete $2I+1$ Hilbert space, so that the $\pi$ pulse is an exact
collective rotation $\exp(-i\pi I_{\varphi_\pi})$. The engineered state (spin
squeezing at $R = 0.7$; an even cat with $|\beta|^2 \approx 12$) forms and holds as
in the main text.

We add a static collective detuning $\delta I_z$. Without refocusing (grey) it destroys the
engineered state: the squeezing is pushed above the standard quantum limit and
the cat fidelity collapses to $\mathcal{F}\approx 0.1$. Interleaving the
single-$\pi$ refocusing train with the frame correction above (cherry) restores
the ideal result, tracking the undisturbed reference (blue dashed).

\section{Optimal engineered dissipation rate}
\label{app:HP_Bogo}
\begin{figure}
    \centering
    \includegraphics[width=\linewidth]{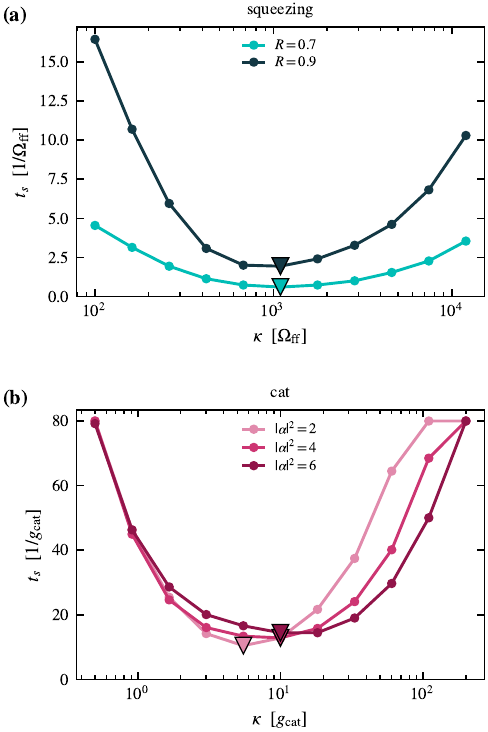}
    \caption{\textbf{Optimal engineered dissipation rate.} Stabilization time $t_s$
    versus the engineered dissipation rate $\kappa$, U-shaped for both protocols: too weak and the adiabatic elimination of the electron breaks down, too strong and the quantum Zeno effect suppresses the coherent interaction; the minimum (markers) is the operating point.
    \textbf{(a)} Squeezing, for several ratios $R=|\Omega_-/\Omega_+|$.
    \textbf{(b)} Cat, for several sizes $|\beta|^2$. Rates in units of
    $\Omega_{\rm ff}$ (squeezing) and $g_{\rm cat}$ (cat); $t_s$ in the inverse of each.}
    \label{fig:stab}
\end{figure}

The engineered dissipation rate $\kappa$, set by the weak-measurement strength,
has an optimum unique to stroboscopic reservoir engineering. The stabilization
time $t_s$ (to reach the steady state within a fixed trace-distance tolerance) is
U-shaped in $\kappa$ for both protocols (Fig.~\ref{fig:stab}): at weak $\kappa$ the adiabatic elimination of the electron is no longer valid, while at strong $\kappa$ the quantum Zeno effect~\cite{Facchi.2002} suppresses the coherent Floquet interaction. The minimum, at intermediate $\kappa$, balances the two and fixes the operating point used throughout.
\section{Effect of ensemble disorder}
\label{app:disorder}

In a realistic quantum dot two sources of inhomogeneity are present: the
hyperfine couplings $g_k$ vary across the nuclear ensemble, and the nuclear
Larmor frequencies $\omega_k$ are not identical.  Both perturb the idealized
collective description of the main text, but through qualitatively different
mechanisms.  Coupling inhomogeneity changes which collective mode the
engineered jump acts on; the error in the bosonic algebra of that mode is
suppressed by the Liouvillian gap of the stabilizer.  Frequency disorder
drives population into sectors of lower total angular momentum $I < N/2$; the
stabilizer continues to act within each sector but cannot restore the lost
angular momentum, and beyond a threshold the stabilized quantum
correlations are degraded.

\subsection{Coupling inhomogeneity}
\label{app:inhomogeneity:coupling}

\subsubsection{Weighted bright mode}

With non-uniform couplings $g_k$ the jump operators are built from
$\tilde I_\pm = \sum_k g_k I^{(k)}_\pm$.  Near the fully polarized state,
writing $I^{(k)}_z = -I_0 + n_k$,
\begin{equation}
  [\tilde I_-,\tilde I_+]
  =- 2\sum_k g_k^2 I^{(k)}_z
  \approx 2I_0\sum_k g_k^2 - 2\sum_k g_k^2 n_k ,
  \label{eq:weighted_comm}
\end{equation}
so $\tilde b \equiv \tilde I_-/(2I_0\sum_k g_k^2)^{1/2}$ obeys
$[\tilde b,\tilde b^\dagger]\approx1$: the ensemble closes as a single
bosonic bright mode, the disorder-weighted one.  The form of the
engineered jump is unchanged within this mode, whether linear
($L_{\rm sq} \propto \Omega_+\tilde I_- + \Omega_-\tilde I_+$) or quadratic
($L_{\rm cat} \propto \tilde I_-^2 - \alpha^2$), so both the squeezed and
cat targets are stabilized by exactly the standard mechanisms.

\subsubsection{Leakage and gap-protected correction}

The correction term $2\sum_k g_k^2 n_k$ in Eq.~\eqref{eq:weighted_comm}
causes $[\tilde b,\tilde b^\dagger]$ to deviate from unity; the relative
size of the deviation is
\begin{equation}
  \eta = \frac{\sum_k g_k^2 n_k}{I_0\sum_k g_k^2}.
  \label{eq:eta}
\end{equation}
In the bright-mode steady state each spin carries occupation
$n_k = c_k^2 n_b$, where $c_k = g_k/(\sum_j g_j^2)^{1/2}$ and
$n_b = \langle\tilde b^\dagger\tilde b\rangle_{\rm ss}$ is the total
bright-mode occupation ($n_b = |\beta|^2$ for the cat).  Substituting
and using $\sum_k c_k^4 \simeq 1/N$ for a Gaussian electron envelope,
\begin{equation}
  \eta \simeq \frac{n_b}{I_0 N}.
  \label{eq:eta_direct}
\end{equation}
The imperfect bosonic closure is a perturbation $\delta\mathcal{L}$ of
magnitude $\Delta_{\rm sq}\eta$ to the Lindbladian.  Because the jump
contains $\tilde I_+$ as well as $\tilde I_-$, it couples back any
population that has leaked into the orthogonal sector, so the perturbed
steady state remains unique.  The Liouvillian gap then gives
$\|\delta\rho_{\rm ss}\|\lesssim\Delta_{\rm sq}\eta/\Delta_L$, and
\begin{equation}
  \Delta\xi^2 \approx \frac{\eta \Delta_{sq}}{\Delta_L}
  = \frac{n_b}{I_0N(1-R^2)},
  \label{eq:xi2_coupling}
\end{equation}
with $\Delta_L = \Delta_{\rm sq}(1-R^2)$ for squeezing; the same $\eta$
enters for the cat with the corresponding gap $\Delta_L^{\rm cat}$.
Numerical values are in Table~\ref{tab:leakage}.

\begin{table}[t]
  \centering
  \begin{tabular}{lcc}
    \hline\hline
    operating point & $n_b$ & $\eta\ (N=2\times10^4)$ \\
    \hline
    squeezing $R=0.7$ & $0.96$ & $1\times10^{-4}$ \\
    squeezing $R=0.8$ & $1.8$  & $2\times10^{-4}$ \\
    squeezing $R=0.9$ & $4.3$  & $4\times10^{-4}$ \\
    cat $|\beta|^2=5.5$ & $5.5$ & $6\times10^{-4}$ \\
    cat $|\beta|^2=11$  & $11$  & $1\times10^{-3}$ \\
    \hline\hline
  \end{tabular}
  \caption{Leakage parameter $\eta$ from Eq.~\eqref{eq:eta_direct}
  (spin-$\tfrac12$ nuclei, $I_0=1/2$, Gaussian envelope, $N=2\times10^4$, so total
  $I=N/2=10^4$).  The correction to the steady-state figure of merit,
  $\Delta\xi^2\approx\eta/(1-R^2)$ [Eq.~\eqref{eq:xi2_coupling}], is of order
  $10^{-3}$ and falls as $1/N$.}
  \label{tab:leakage}
\end{table}

\subsection{Frequency disorder}
\label{app:inhomogeneity:freq}

Frequency disorder, a spread of the individual nuclear Larmor frequencies, dephases the
nuclei~\cite{Urbaszek.2013}; we model it as independent local dephasing at rate $\gamma$
with jump operators $\sqrt{\gamma}\sigma_z^{(k)}$.

These operators break permutation symmetry, leaking population from the $I=N/2$
sector into sectors of lower total angular momentum $I<N/2$ at rate $\sim\gamma$, which
we simulate exactly in the permutation-invariant Dicke basis using the PIQS
library~\cite{Shammah.2018}.
Crucially, this does not switch off the stabilization: the collective jump operators
act \emph{within} every sector, so the engineered dissipation continuously re-forms
its dark state, a squeezed (or cat) state of whatever $I$ the population currently
occupies. Leakage therefore does not destroy the target; it only lowers the total
angular momentum $I$ it is built on. Because the SQL and fidelity reference is fixed to
$N/2$, the metric responds only to this reduction of $I$; at large $I$ each step is a
vanishing fractional change, so the state degrades only after many leakage events.

\begin{figure}
  \centering
  \includegraphics[width=\columnwidth]{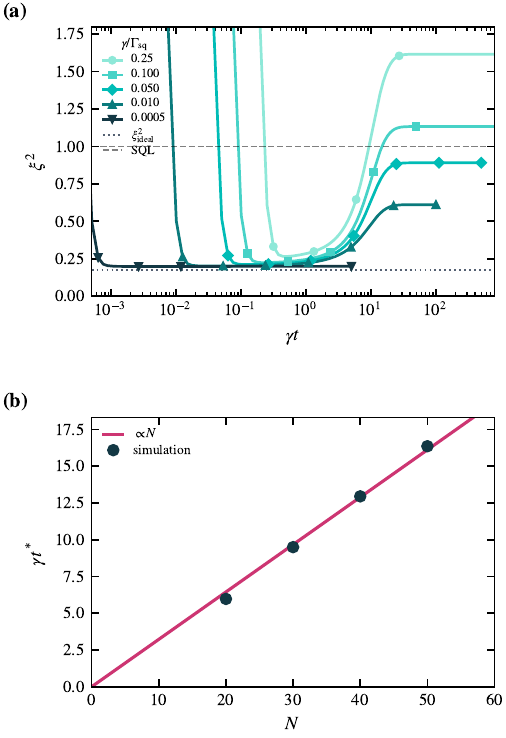}
  \caption{%
    Local dephasing robustness.
    \textbf{(a)}~Squeezing parameter $\xi^2(t)$ for five disorder ratios
    $\gamma/\Gamma_{\rm sq}$ (dephasing rate $\gamma$, engineered rate $\Gamma_{\rm sq}$;
    $N=20$, $R=0.7$).  Three regimes are visible:
    formation on the $\Gamma_{\rm sq}^{-1}$ timescale, leakage-driven rise
    beginning at $\gamma t\sim1$, and a steady-state plateau whose level
    depends on $\gamma/\Gamma_{\rm sq}$.  SQL ($\xi^2=1$, dashed) and
    $\gamma=0$ ideal limit (dotted) shown for reference.
    \textbf{(b)}~Degradation timescale $\gamma\tau^*$ versus $N$ ($N\geq20$); the points follow
    the linear scaling $\gamma\tau^*\propto N$ (line).
  }
  \label{fig:fig_app_4_dephasing}
\end{figure}

Three timescales are visible in Fig.~\ref{fig:fig_app_4_dephasing}(a): the target forms
fast, on the engineered timescale $\Gamma_{\rm sq}^{-1}$; leakage sets in later, at
$\gamma t\sim1$; and the steady-state plateau, set by the ratio $\gamma/\Gamma_{\rm sq}$
of the dephasing rate to the engineered rate, stays below the SQL for weak disorder.

This is why larger ensembles are more robust: draining
the total angular momentum from $I=N/2$ to the intra-sector breakdown threshold requires
$O(N)$ successive leakage steps, each at rate $\sim\gamma$ and each a vanishing
fractional perturbation at large $I$, so $\gamma\tau^*\propto N$. The simulation confirms
this linear scaling [Fig.~\ref{fig:fig_app_4_dephasing}(b)]. The degradation time is
therefore $\tau^*\sim N/\gamma$ (up to an $O(1)$ factor); with $\gamma=2/T_2$ and the
register size $N\sim2\times10^4$, this is of order $10$s at $T_2\approx1$~ms and
$\sim10^3$s once dynamical decoupling reaches $T_2\approx100$~ms, in both cases well
beyond the storage window used here.

The cat follows the same leakage mechanism, and its parity is untouched: like
collective dephasing, local dephasing $\sqrt{\gamma}\sigma_z^{(k)}$ is diagonal in
$I_z$ and conserves $e^{i\pi I_z}$, so it cannot drive a logical even-odd flip.
Frequency disorder therefore degrades the cat only through the same slow
angular-momentum drift as the squeezed state, with its parity protection intact.

\section{Numerical methods}
\label{app:numerics}

All figures come from direct simulation of the stroboscopic cycle at the pulse
level with no adiabatic elimination assumed, except for the effective
master-equation reference curves in Fig.~\ref{fig:ideal}. Here we give only the
numerical specifics: how the cycle is represented and iterated, and how each panel is
read off.

\subsection{Representation and dephasing}

The joint electron-nuclear state is propagated as a dense density matrix on
$\mathbb C^2\otimes\mathbb C^{N+1}$, keeping the top $N+1$ Dicke states
$|I,I-n\rangle$; the depth $N$ is fixed by a spread margin $k_\sigma$. One cycle
applies the coherent pulse evolution followed by the weak measurement, with the per-period composite unitaries precomputed so
that a cycle is a few dense $2(N{+}1)$ matrix multiplications. Collective dephasing is
added as a per-cycle damping of each coherence $\rho_{ab}$ by
$\exp[-\tfrac12\gamma t_c(z_a-z_b)^2]$ in the $I_z$ eigenbasis, exact when the per-cycle
dephasing is small, $\gamma t_c\ll1$, and quoted as the single-quantum coherence time
$T_2=2/\gamma_{\rm phys}$. Throughout we fix the ensemble size $I=10^4$, the collinear
(secular) hyperfine coupling $A_c=0.11\%\omega_n$, and the per-cycle
drives the engineered interaction is the parameter varied between the two targets and
weak-measurement strength $\epsilon$ is set at the stabilization optimum, with $\epsilon\leq0.1$; the non-collinear coupling $A_{\rm nc}$ that
swept in Fig.~\ref{fig:noise}. The simulation runs with $\omega_n=1$; the physical
anchor $\omega_n/2\pi=100$~MHz fixes the cycle time in seconds and the $T_2$ axis of
Fig.~\ref{fig:noise}.

\subsection{Figure~\ref{fig:ideal}: ideal time traces ($\gamma=0$)}

\emph{(b) Squeezing.} The cycle map is iterated from the polarized pole
($k_\sigma=22$; cycle time $t_c=170$ and $190$~ns for $R=0.7$ and $0.9$), reading the
Wineland parameter $\xi^2_R$ of the reduced nuclear state at logarithmically spaced
snapshots. Both $R=0.7$ and $R=0.9$ are iterated directly (to $1$~s and $\sim0.19$~s; floors $-7.53$ and $-12.8$~dB). The faint reference
curves solve the joint master equation of Appendix~A,
$\dot\rho=-i[H_{\rm int},\rho]+\kappa\mathcal D[S_-]\rho$ with
$H_{\rm int}=\Omega_{\rm fl}(S_+I_++S_-I_-)+\Omega_{\rm ff}(S_+I_-+S_-I_+)$, from the
polarized pole with the working-point couplings $\Omega_{\rm ff},\Omega_{\rm fl},\kappa$
derived in the text.

\emph{(c) Cat.} The skip cycle [Appendix~B] is iterated from the polarized state,
integrating the continuous drive in $n_\varphi=4920$ time steps per block; $\mathcal F$
is the fidelity to the converged even cat. The densely sampled formation ramp is
continued to the $1$~s plateau by exact repeated squaring of the one-cycle stroboscopic
propagator $\Phi^{2^k}$, which reaches $10^8$ cycles without accumulating per-step
integration error. The faint curves solve the corresponding two-photon master equation,
$\dot\rho=-i[\chi(S_+I_-^2+S_-I_+^2)+\Omega S_x,\rho]+\kappa\mathcal D[S_-]\rho$, with
$\chi$ from the van Vleck Fourier sum of Appendix~B and $\kappa=\epsilon/t_c$; this
parity-exact reference is drawn over the formation ramp, where it coincides with the
full dynamics before the higher-order plateau.
\subsection{Figure~\ref{fig:noise}: operating-region grids}

At each coupling $A_{\rm nc}$ and dephasing rate $\gamma$ we run the full dephased
stroboscopic cycle and read out the target's quality parameter. The per-cycle measurement
is set at the stabilization optimum (App.~\ref{app:HP_Bogo}): its rate scales with the
collective coupling and is capped at a per-cycle strength $\epsilon\le0.1$. Both grids are
logarithmic in $A_{\rm nc}/\omega_n$ (over $\sim10^{-4}$--$3\times10^{-3}$) and in
$T_2=2/\gamma$ ($\sim10\mu$s to $\gtrsim1$~s).

\emph{(a) Squeezing ($R=0.90$).} The squeezed state is the steady state, so we iterate the
dephased cycle to convergence ($|\Delta\xi^2_R|<0.04$~dB) and record $\xi^2_R$ (truncation
$k_\sigma=18$, $\dim_n=126$). Contours mark the SQL ($\xi^2_R=1$) and $-6$~dB.

\emph{(b) Cat ($|\beta|^2\approx5.5$, $N_{\max}=24$).} We evaluate the fidelity to the
even cat after $1$~s of storage by exact repeated squaring of the one-cycle dephased
superoperator ($\Phi^{2^k}$, reaching the $\sim5\times10^6$ cycles of $1$~s). Contours mark
$\mathcal F=0.99$ and $0.90$.
\subsection{Convergence}

The squeezing steady state requires $k_\sigma\gtrsim18$: at shallower truncation a
spurious rebound of $\xi^2_R$ appears after formation (a top-of-ladder truncation
artifact) that both corrupts the value and defeats the convergence test. The cat
requires the skip cycle and a converged drive discretization ($n_\varphi\approx4920$)
to dynamically decouple $A_{\rm nc}$; at coarser $n_\varphi$ a spurious single-quantum
leak survives. 

\end{document}